\documentclass{elsart}
\usepackage{graphicx}
\usepackage{amssymb}

\begin{document}

\begin{frontmatter}
\title{Monte Carlo Simulations for the Slow Relaxation of Crumpled Surfaces}
\author[Uff,Unipli]{Klauko Mota}\ead{klauko@if.uff.br}
\author[Uff]{and Paulo Murilo C. de Oliveira}
\address[Uff]{Universidade Federal Fluminense, Instituto de F\'isica \\Av. Litoran\^ea s/n Boa Viagem Niter\'{o}i, Rio de Janeiro, Brazil, 24020-005}
\address[Unipli]{ Centro Universit\'ario Pl\'inio Leite, Departamento de Ci\^{e}ncia da Computa\c{c}\~{a}o \\Av. Visconde do Rio Branco 123, Centro, Niter\'oi, Rio de Janeiro, Brazil}

\begin{abstract}

In this paper we study crumpled surfaces through Monte Carlo Simulations. The crumpled surface is represented by a cluster of spins  pointing up and spins pointing down represent the air both inside and around the surface. We follow the time dynamics of this fractal 
structure and we have shown that it presents a stretched exponential behaviour.
\end{abstract}

\end{frontmatter}
\section{Introduction}
Many systems have slow relaxation, and a wide variety of them are well described by the stretched exponential form
\begin{equation}
f(t) = f_0\exp(-(t/\tau)^{\beta}),
\end{equation}
with the stretching index $\beta$ ranging from $0$ to $1$, where the borderline $\beta=1$ corresponds to the usual exponential function. The stretched function was firstly suggested by Kohlrausch \cite{1} in the middle of the 19th century to study viscoelasticity and after postulated by Williamns and Watts \cite{2} to analyse dielectric relaxation. Often, this function is associated with structural relaxation in glasses and polymers \cite{2,3} but recently it is known also for spin and magnetic glasses \cite{4,5}, random magnets and economic behaviour.

In particular, an interesting physical phenomenon which also shows slow relaxation is the process of crumpling. This phenomena is poorly understood and it can be associated with several situations in nature. For instance, the way energy is dissipated in a crashing (ie, a car crash) is an important subject of investigation.

A convenient and, of course, economical means of studying this phenomena in laboratory is the crumpling of sheets of different materials such as aluminum foils, Mylar sheets and paper \cite{6}. For example, a sheet of paper can be crumpled, squeezed by hands into a ball as tightly as it is possible. The resulting object has fractal structure that satisfy the mass-size scaling relation  $ M = R^D$, where $D$ is around $2.5$ \cite{7,8}.

\begin{figure}[ht]
\begin{center}
\includegraphics[height=3.0cm]{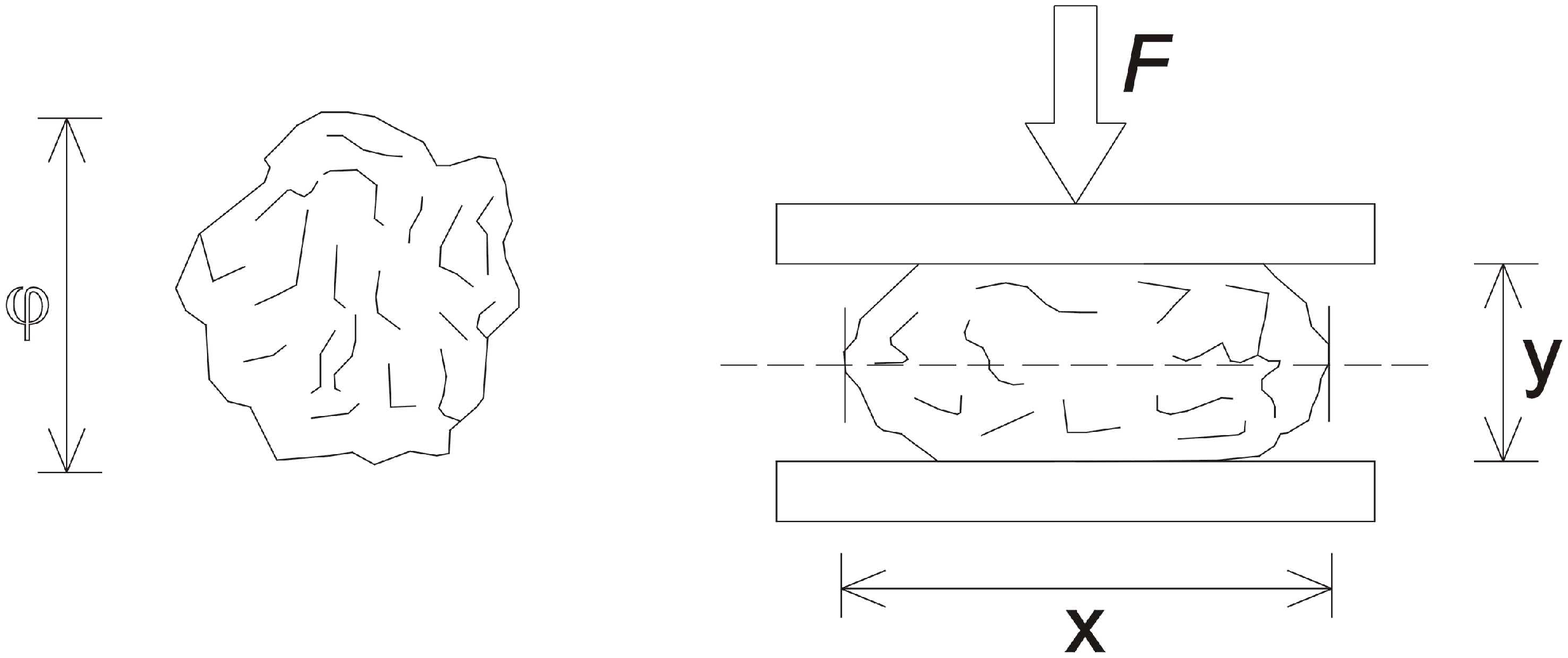}
\caption{The crumpled surface process used in the Albuquerque and Gomes' experiment \cite{9}.}
\label{fig1}
\end{center}
\end{figure}

Albuquerque and Gomes studied stress relaxation in crumpled surfaces experimentally \cite{9}. The crumpled surface in their experiment was obtained from aluminum foils carefully crumpled by hands in an approximately spherical shape. The crumpled surface is then inserted between two parallel plates and compressed by an axial force during a time $\Delta t$, needed to establish a small deformation. After such time, the plates were kept fixed and the stress decay was then measured on the plates. Figure \ref{fig1} illustrates this process in a schematic manner.

The experiments of Albuquerque and Gomes showed a slow relaxation of the mechanical stress. Their results indicate an anomalous stretched exponential behaviour for the stress decay. They believe that this behaviour is possibly a direct consequence of the fractal structure of the crumpled surface.

The aim of this paper is to show that there is a relationship between slow relaxation and fractal structure. To do that, we consider a simple model based on Ising spins to capture this interesting behaviour. This model, known in the literature as dynamic drop model, was applied in several out-of-equilibrium systems such as multifragmentation in both nuclear matter and mercury drops falling on the ground, leaky faucets, magnetic hysteresis, etc \cite{10,11}.

We do not pretend to describe the many features observed in a crumpled surface, a very complex system, with our simple model. Instead, its focus is only to relate the slow relaxation with the fractal internal structure, according to the hyphotesis raised by Albuquerque and Gomes.

\section{The Model}

The crumpled surface in our simulations is represented by a cluster of Ising spins pointing up ($S_i = +1$) on a lattice. Spins pointing down ($S_i = -1$) represent the air that can be both inside and around the crumpled surface. Spins pointing up are distributed in such a way that they form a fractal picture. We can build for example a regular or a statistical fractal from these spins, just by implementing a dynamic rule. In our simulations we consider the Sierpinski carpet and the Sierpinski gasket as examples of regular fractal and the percolation cluster as example of statistical fractal.

To take into account a fixed strain on the crumpled surface, we consider two parallel external boundary lines of spins, the plates, one on the top and the other on the bottom of the crumpled surface. These spins are kept fixed during all the time evolution. The ferromagnetic interaction between the spins and these fixed boundary keeps the surface glued.

To avoid the fragmentation of the surface, the spins are coupled through a first-and-second neighbour ferromagnetic interaction, with the following Hamiltonian
\begin{equation}
H = -\sum_{<ij>}S_i S_j,
\label{eq1}
\end{equation}
where the sum runs over all pairs $<ij>$ of spins which are first or second neighbours on the square lattice.

This interaction can be interpreted otherwise. Consider null the attractive energy stored in each pair of interacting spins pointing 
parallel to each other. The whole energy is then stored on the up-down interface, therefore distributed along the cluster perimeter - see equation (\ref{eq2}) below. The Ising Hamiltonian, equation (\ref{eq1}), is also a model for surface tension energy.

A Monte Carlo updating of the spins is then performed. The flipping of the spins is accepted through the Glauber algorithm following a 
Kawasaki-like dynamics, with conservation of mass (number of up or down spins constant) \cite{12}. Each updating step consists in choosing two random spins, one pointing up and the other pointing down. The spin pointing up belongs to the current inner boundary of the spin-up cluster. The other spin pointing down belongs to the current outer boundary. Then, both are flipped if this movement decreases the total energy given by equation (\ref{eq1}), respecting external boundary conditions.

\section{Results}

The pressure exerted by the surface on the plates \cite{9} can be defined within our model as the number of up spins that are in contact with the plates, i.e., the external boundary line of spins. For example, a surface in form of a square of side $L$ initially would have pressure equal to $2L$. As we are using the multispin technique, we can easily compute this pressure with a simple boolean instruction.

In the case of the energy relaxation, on the other hand, we can use a trick commonly used in Monte Carlo simulations. As the relaxation process always tends to diminish the total energy of the system, we can initially keep in the computer memory the initial value of the total energy and, whenever it decreases after one movement, we update the energy, i.e.
\begin{equation}
    E(t+1) = E(t) - \Delta E.
\end{equation}

The energy as a function of time is associated with the crumpled surface perimeter. The ferromagnetic coupling represented by equation (\ref{eq2}) keeps the surface connected into a single macroscopic piece, with holes inside. Indeed, it could be written as
\begin{equation}
H = 2\sum(1-\delta_{S_i,Sj}),
\label{eq2}
\end{equation}
plus an additive irrelevant constant. Under this point of view, only the surface boundary contributes to the energy. In fact, only at the surface boundary one can find two neighbouring spins $S_i$ and $S_j$, one pointing up inside the surface, the other pointing down outside, contributing to the above sum. Otherwise, far from the boundary, one has $S_i=S_j$ (both neighbouring spins pointing up inside the surface or both down outside), with no contribution to the above sum.

For the data analysis, we used a free code package called Gnuplot. This package, in turn, uses internally an implementation of the nonlinear least-squares Marquardt-Levenberg algorithm \cite{13} for curve adjustments. The quality of the adjustments can also be evaluated by this package. After each iteration, the package shows information that can be used to analyse the evolution of the adjustment. For example, to each step of iteration the package calculates the value of $\chi^2$ per degree of freedom and the matrix of correlation of the parameters used for the adjustment.  For a good fit, $\chi^2$ per degree of freedom (also reduced $\chi^2$) is near unity.  On one hand, if it is much bigger, ones underestimated the error bars or the fitting function may not describe the data very well. On the other hand, if it is much smaller, ones overestimated the error bars.

The fitting function used in our data analysis is the stretched exponential $f(t) = f_1\exp(-(t/\tau)^\beta) + f_2$, where $f$ could be 
the energy or the stress and $f_1$, $f_2$, $\tau$ and $\beta$ are constant parameters that will be found in the fit process. The usual 
exponential decay corresponds to the same $f(t)$ with $\beta = 1$.

Our results presented later (figures 3, 4, 7, 8, 11, 12, 15 and 16) are obtained after simulating many samples of the same system (hundreds), in order to get very small error bars (hardly visible in the plots). This makes the computational effort also hard. On the other hand, it allows us to obtain the exponent $\beta$ within the accuracy shown in the figure captions (three digits, with the last one uncertain).

In this section we show that the crumpled surfaces used in our simulations present slow relaxation ($\beta < 1$) when the lattice is 
fractal. In the next section, we find faster relaxation ($\beta = 1$) whithout fractality. Indeed, the same behaviour hyphotetically raised by Albuquerque and Gomes.

We do not pretend to make quantitative comparisons of the $\beta$ values with reality. That is why we decide to simulate 2D instead of 3D systems, otherwise the computational effort would become prohibitive. Moreover, one cannot be sure if the precise value of $\beta < 1$ is influenced by other features (as friction due to two touching surfaces, their self-avoiding character, etc), which are absent from our model.

\subsection{Percolation Cluster}

An example of statistical fractal is the percolation cluster. In our simulations it is generated dynamically, starting from a single point 
(the origin), via the Leath algorithm \cite{14}. Initially, all sites on the lattice are considered virgin sites, except the origin. Then, 
recursively, the neighbours of the neighbours of the neighbours... of the initial site are occupied with probability $p$ and otherwise 
unoccupied. In both cases, the site is considered no longer virgin, and is not tested again. This process continues until the cluster stops growing (due to the formation of a boundary of unoccupied sites) or until an outer boundary of the system (e.g.  our two plates) is reached.

\begin{figure}[ht]
\begin{center}
\includegraphics[height=6cm]{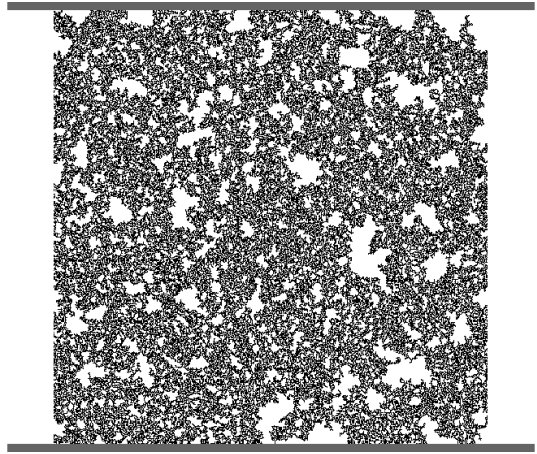}
\caption{An example of percolation cluster used as initial lattice for our dynamic evolution.}
\label{fig2}
\end{center}
\end{figure}

As we are interested in a fractal cluster, one generates it with probability $p=p_c$, where $p_c$ is the percolation threshold. The most important advantage in using the Leath algorithm is that we generate only the spanning cluster, which essentially is a fractal and its fractal dimension is about $d_f \approx 1.8958$ in 2D. Figure (\ref{fig2}) shows a spanning cluster in a lattice of linear size $L = 
512$. Gray lines both in bottom and top represent the external boundary condition that forbid the cluster from growing indefinitely.

\begin{figure}[ht]
\begin{center}
\includegraphics[height=7.5cm]{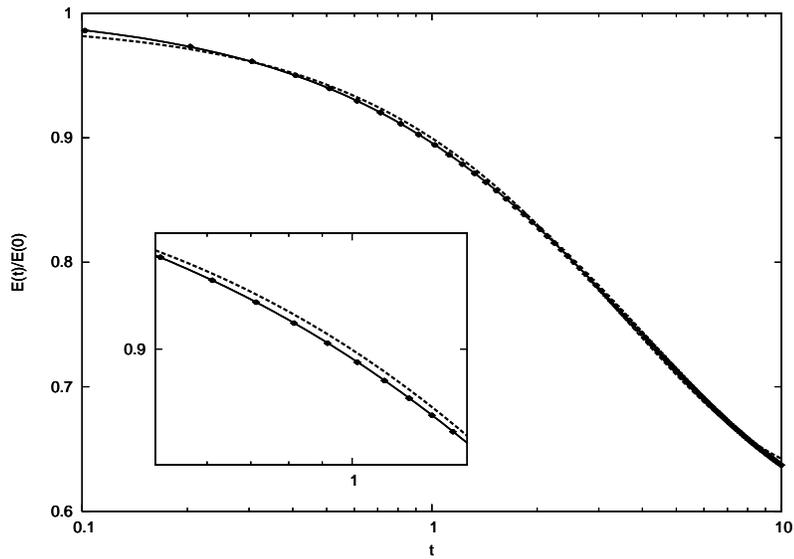}
\caption{Energy as function of time.  Full line is the stretched exponential fit with reduced $\chi^2 = 1.16$, the stretched exponent is 
$\beta = 0.874$. Dashed line is the exponential fit, where the reduced $\chi^2 = 704.70$.}
\label{fig3}
\end{center}
\end{figure}

\begin{figure}[ht]
\begin{center}
\includegraphics[height=7.5cm]{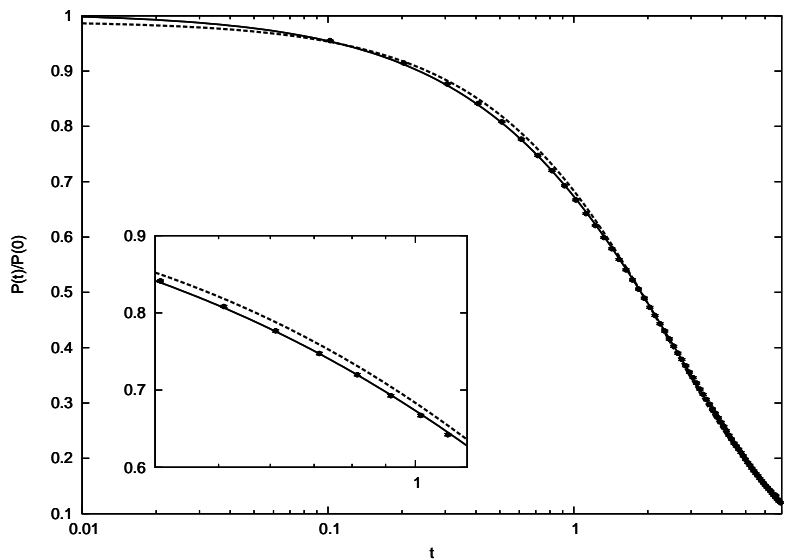}
\caption{Stress as function of time. Full line is the stretched exponential fit with reduced $\chi^2 = 0.99$, and the stretched exponent 
is 0.893. Dashed line is the exponential fit, where the reduced $ \chi^2 = 23.02$.}
\label{fig4}
\end{center}
\end{figure}

In our simulations, we consider spanning clusters generated in a square lattice of size $L=1024$. Our results indicate an exponential stretched behaviour both in energy as in stress relaxation. The best fits are shown in figures (\ref{fig3}) and (\ref{fig4}). In order to show the differences between stretched and normal exponential behaviour, we also plot in the same figure a fit using an exponential function as a possible model to the crumpled surface relaxation.

\begin{figure}[ht]
\begin{center}
\includegraphics[height=12cm]{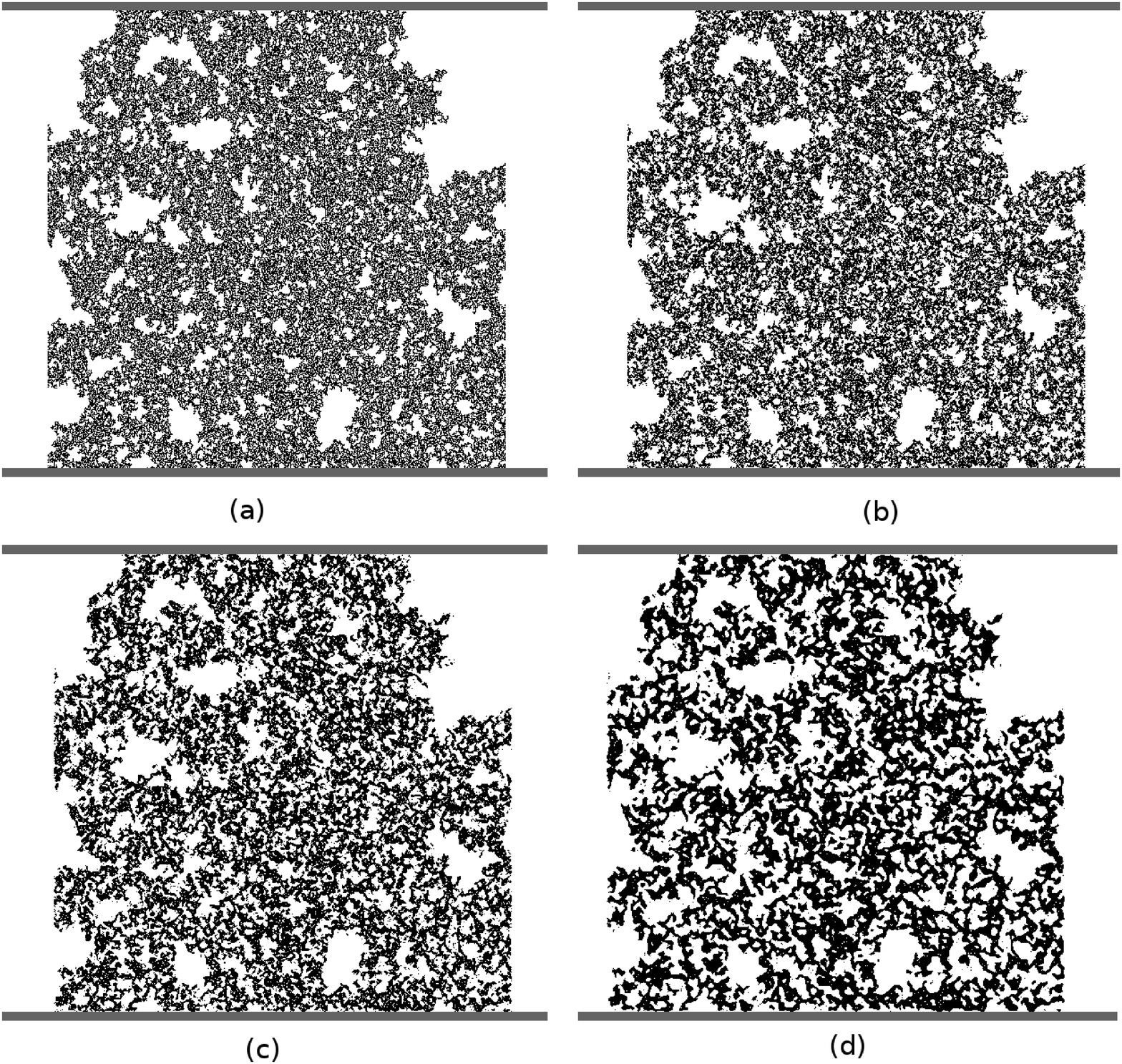}
\caption{Time evolution of the starting percolation cluster with mass $M = 114375$ (number of black pixels). (a) Initial lattice created from Leath algorithm, (b) lattice after $2M$ Monte Carlo steps, (c) after $4M$ Monte Carlo steps and (d) after $8M$ Monte Carlo steps. The dynamics runs according to the hierarchy of holes, as follows. First, the smallest holes (isolated white pixels) disappear. Then, holes a little bit larger than one pixel also disappear. Then, larger holes disappear, and so on. The dynamics is defined by the previously posed fractal properties. The same behaviour is seen in figures 9 and 13.}
\label{fig5}
\end{center}
\end{figure}

In figure (\ref{fig5}) we show some steps of the spanning cluster relaxation. Each black pixel represent the crumpled surface (i.e. spin 
pointing up) and white pixel the air (spin pointing down).

\subsection{Sierpinski Carpet}

\begin{figure}[ht]
\begin{center}
\includegraphics[height=7cm]{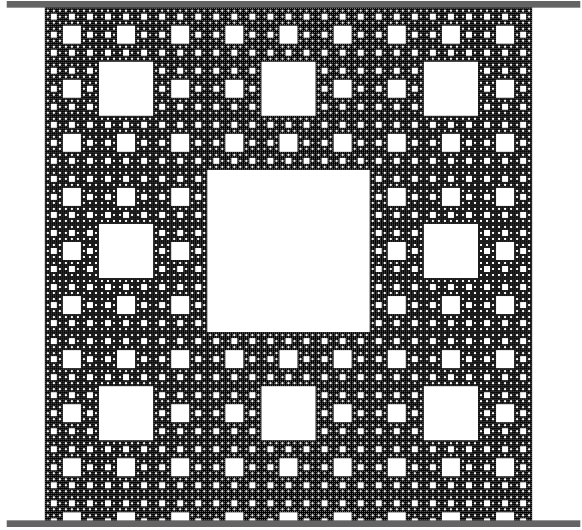}
\caption{Sierpinski carpet used as starting configuration for our dynamic evolution.}
\label{fig6}
\end{center}
\end{figure}

The Sierpinski carpet is a plane regular fractal with fractal dimension approximately equal to $1.8928$ in 2D. The construction of this fractal begins with a solid filled square. Then we divide it into $9$ smaller congruent squares and we remove the central one. The same process is applied recursively to each remaining solid square, \textit{ad infinitum}.  This fractal will be used as an example of crumpled surface in our simulations, see Figure (\ref{fig6}).

\begin{figure}[ht]
\begin{center}
\includegraphics[height=7.5cm]{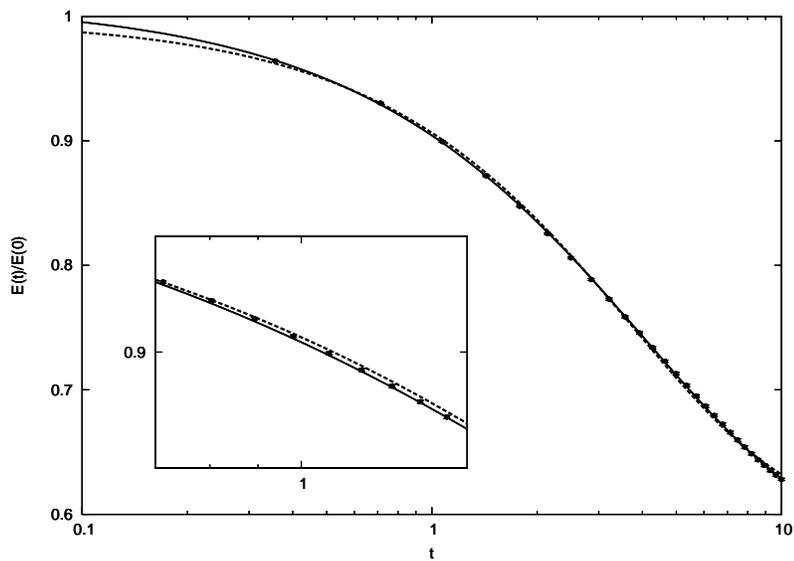}
\caption{Energy relaxation of the Sierpinski carpet. Full line is the stretched exponential fit, where the reduced $\chi^2 = 1.07$ and the 
stretched exponent is $\beta = 0.886$. Dashed line is the exponential fit, where $\chi^2 = 7.05$.}
\label{fig7}
\end{center}
\end{figure}

\begin{figure}[ht]
\begin{center}
\includegraphics[height=7.5cm]{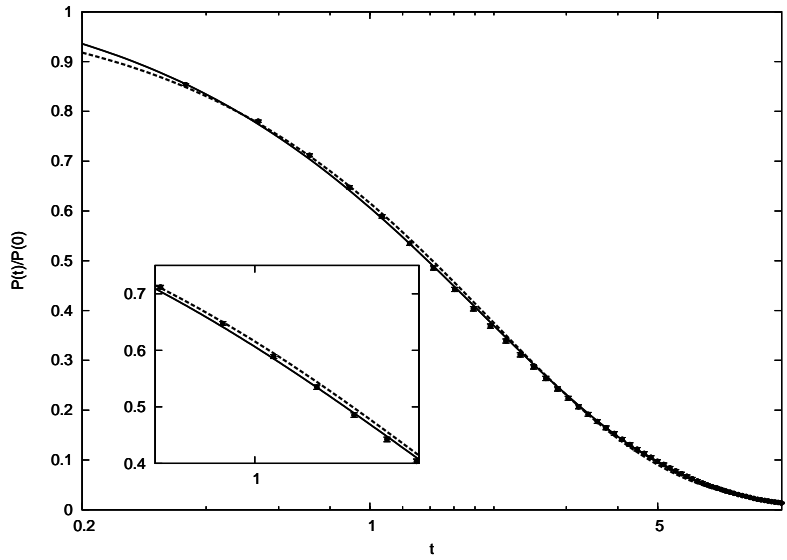}
\caption{Stress relaxation of the Sierpinski carpet. Full line is the stretched exponential fit, where the reduced $\chi^2 = 1.10$ and the 
stretched exponent is $\beta = 0.910$. Dashed line is the exponential fit, where $\chi^2 = 3.35$.}
\label{fig8}
\end{center}
\end{figure}

\begin{figure}[ht]
\begin{center}
\includegraphics[height=10.5cm]{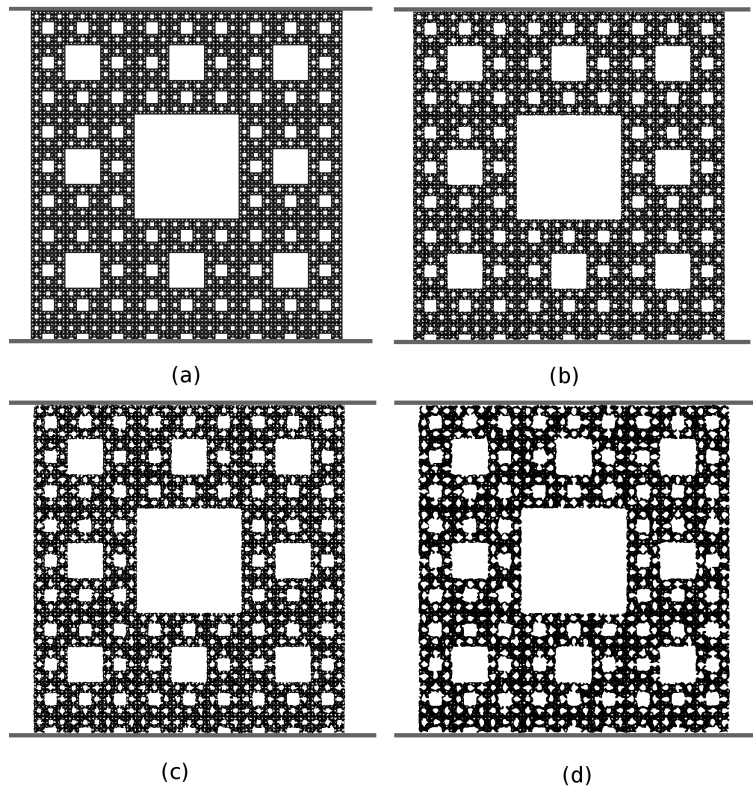}
\caption{Some snaptshots of evolution from Sierpinski Carpet. }
\label{fig9}
\end{center}
\end{figure}

In our simulations we consider the sixth generation of the Sierpinski carpet in a lattice of size $L = 736$. Our results also indicate an 
apparent stretched exponential behaviour, where the stretched exponent $\beta$ is very close to the $\beta$ exponent obtained from percolation cluster. We believe this occurred because both fractal lattices have fractal dimension very close to each other. Again one plots in the same figure the stretched exponential fit and the exponential fit, energy relaxation in figure (\ref{fig7}) and stress relaxation in figure (\ref{fig8}). Figure (\ref{fig9}) shows some configurations during the time evolution.

\subsection{Sierpinski Gasket}

The Sierpinski gasket is another interesting and well-known regular fractal. Its fractal dimension is about $1.585$. It is quite simple to 
generate it, as follows: we start with a solid equilateral triangle. Then, we divide it into four smaller equilateral triangles using the 
midpoints of the three sides of the original triangle as the new vertices. By removing the interiour of the central triangle (not 
removing the outer boundary) we get the first generation. Now, repeating this procedure on each of the three remaining solid equilateral triangles we obtain the second generation and so on.

\begin{figure}[ht]
\begin{center}
\includegraphics[height=7.0cm]{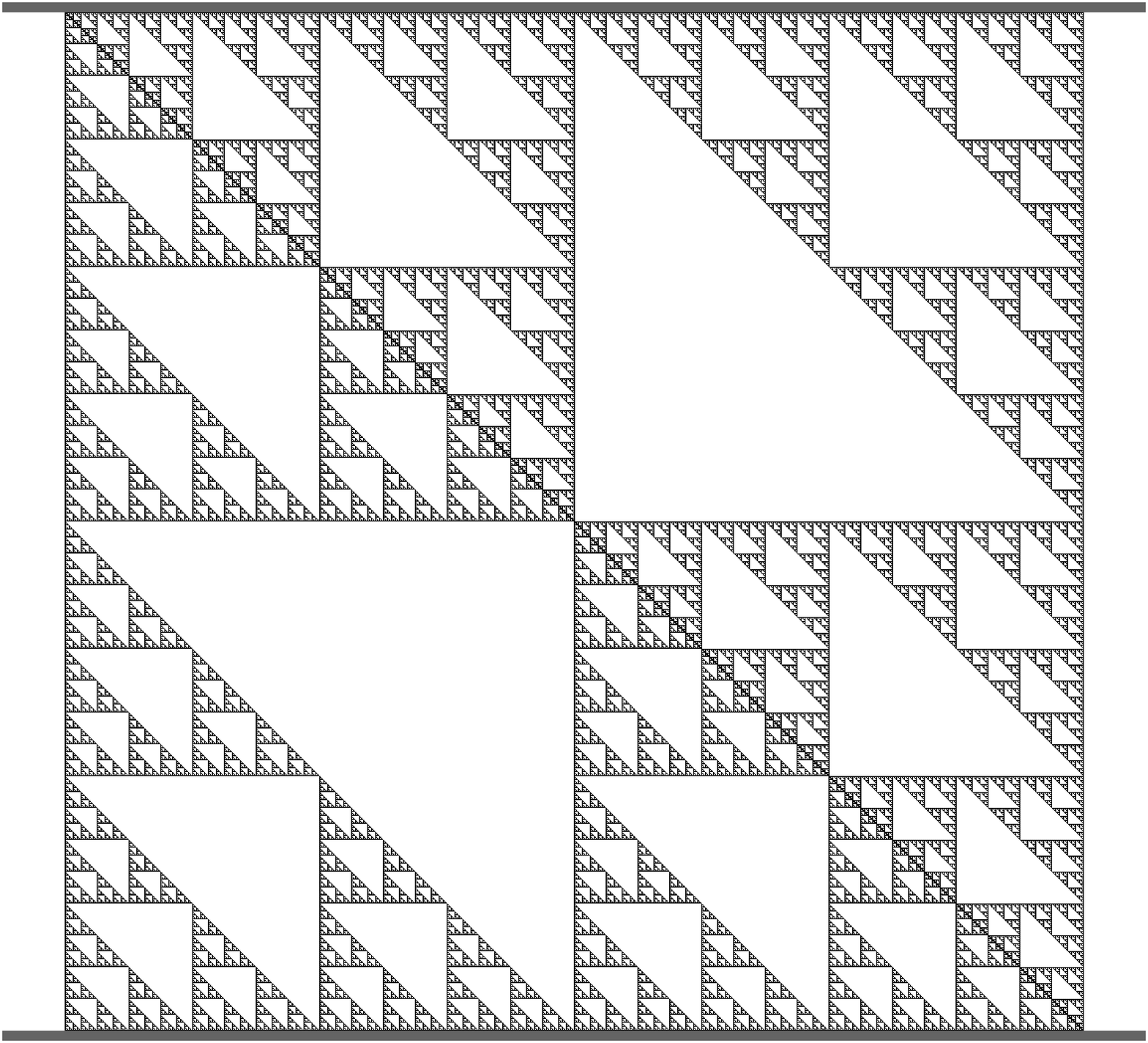}
\caption{Fractal used in our simulations as the starting crumpled surface. It was generated from two Sierpinski gasket in a square lattice. }
\label{fig10}
\end{center}
\end{figure}

Here we consider a different construction for the Sierpinski gasket. Rather than starting with an equilateral triangle, we start with a right triangle with two of its sides of length $L$. The rest of the construction proceeds as before.  In order to get a fractal structure on 
a square lattice, we used two Sierpinski gaskets, as showed in figure (\ref{fig10}).

\begin{figure}[ht]
\begin{center}
\includegraphics[height=7.5cm]{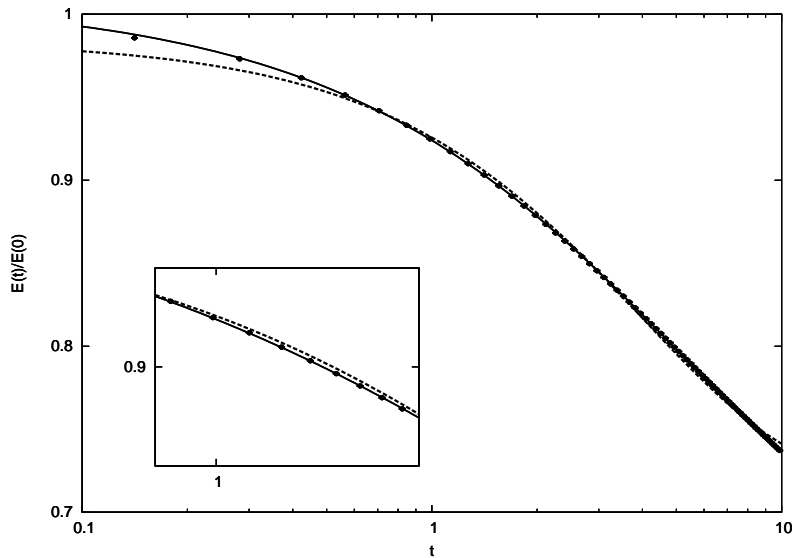}
\caption{Energy relaxation in a gasket-like fractal. Full line is the stretched exponential fit, where the reduced $\chi^2 = 1.06734$ and the stretched exponent is $\beta = 0.758$. Dashed line is the exponential fit, where the reduced $\chi^2 = 102.122$.}
\label{fig11}
\end{center}
\end{figure}

Figure (\ref{fig11}) shows the energy relaxation of the gasket-like fractal. Continuous line represents the fit that was reached by using 
the stretched exponential fitting function. In the same figure ones plots another fit using the exponential function as model. Clearly it 
can be noticed that the stretched exponential model represents the best fit.

\begin{figure}[ht]
\begin{center}
\includegraphics[height=7.5cm]{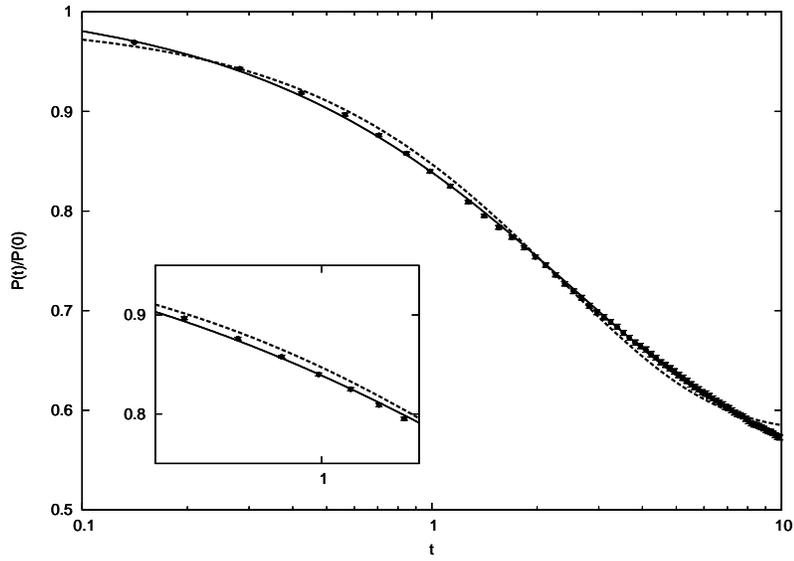}
\caption{Stress relaxation in a gasket-like fractal. Full line is the stretched exponential fit, where the reduced $\chi^2 = 1.06318$ and the stretched exponent is $\beta = 0.794$. Dashed line is the exponential fit, where the reduced $\chi^2 = 10.4779$.}
\label{fig12}
\end{center}
\end{figure}

Figure (\ref{fig12}) shows the stress relaxation to the gasket-like fractal. We plot again in the same plot a fit using the stretched 
exponential function and the exponential function. The best fit was also reached by the stretched exponential function.

\begin{figure}[ht]
\begin{center}
\includegraphics[height=10cm]{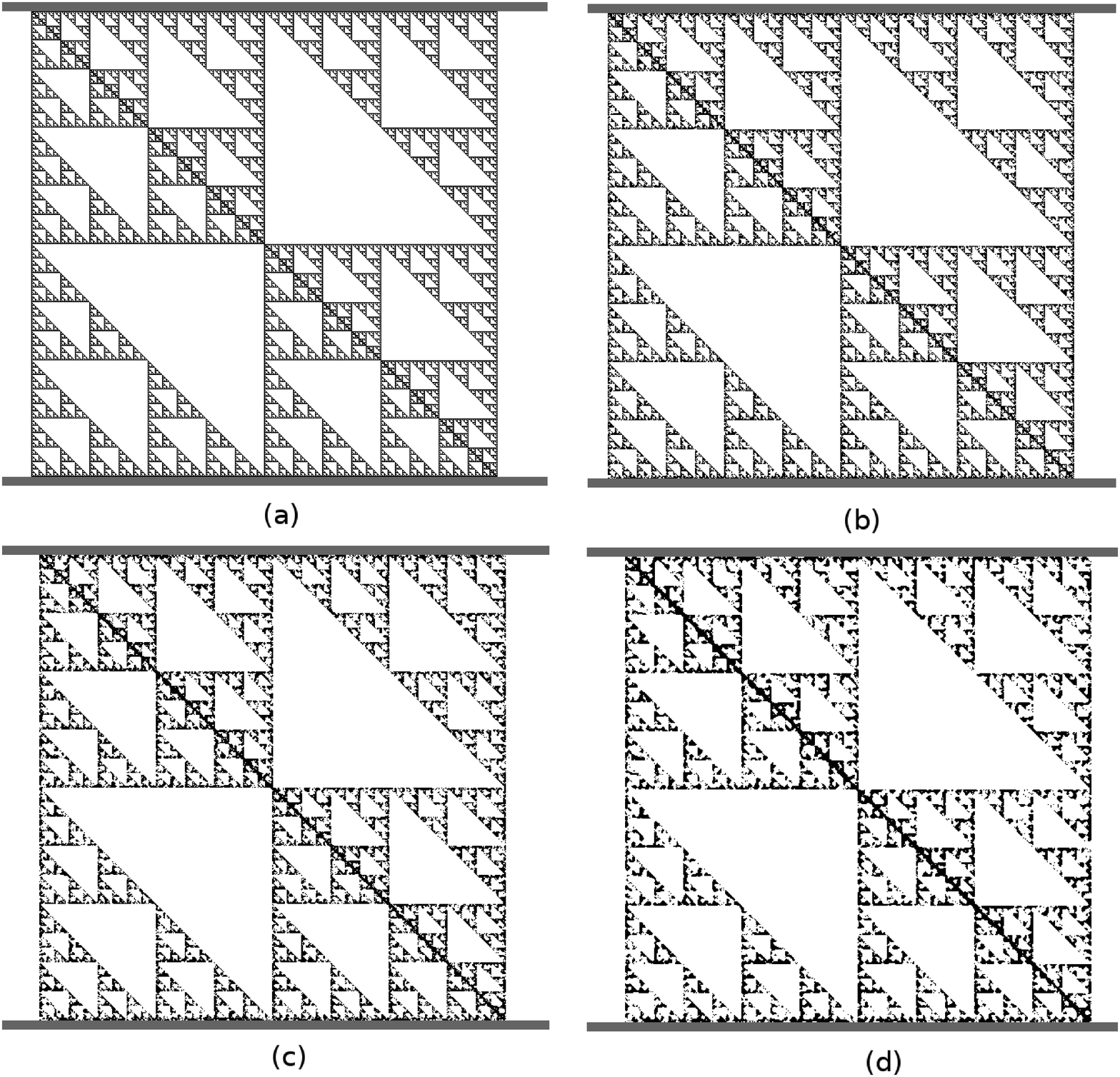}
\caption{Time evolution from a gasket-like fractal with mass $M = 39366$ in a square lattice of size $L = 512$. (a) Initial lattice, (b) lattice after $t = 2M$ steps, (c) lattice after $t = 4M$ steps and (d) lattice after $t = 8M$ steps.}
\label{fig13}
\end{center}
\end{figure}

Figure (\ref{fig13}) shows some snapshots from the gasket-like evolution. As time goes by, the surrounding lines of the gasket-like 
becomes more and more rounded. If we let the system evolve for a long time, the picture becomes more compact, since their holes become filled.

\section{Non-fractal Crumpled Surface}

Up to now we considered in our simulations two regular fractals and one statistical fractal as examples of crumpled surfaces. We have shown that both the system energy and the stress exerted by the crumpled surface on the plates have slow relaxation, following a well-defined stretched exponential form. In fact we believe that such behaviour is related to the fractal characteristics of those lattices. In order to finish this work, we will now simulate a non-fractal surface.

\begin{figure}[ht]
\begin{center}
\includegraphics[height=4.0cm]{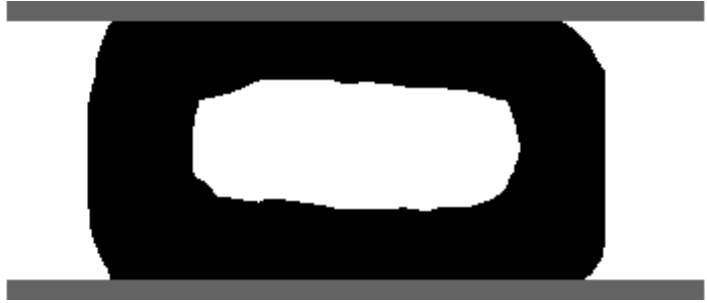}
\caption{Compact crumpled surface created from a graph editor.}
\label{fig14}
\end{center}
\end{figure}

Such surface can be created by many ways, for example, we could simulate a compact surface in the form of a square of side $L$. However, due to the artificial square mesh, the surface would not relax easily, since the exchanging of spins would take too long to happen or would even be almost impossible. Instead, by using a graphical editor, we drew a compact surface that could be relaxed more easily, as can be viewed in figure (\ref{fig14}).

\begin{figure}[ht]
\begin{center}
\includegraphics[height=8.0cm]{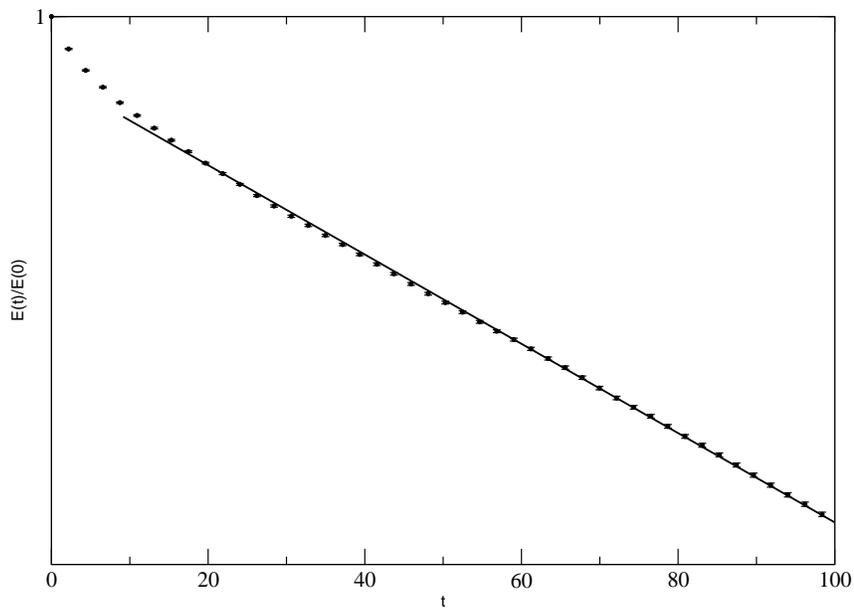}
\caption{Energy relaxation as function of time for the non-fractal initial cluster of figure (\ref{fig14}). After a transient, data are 
well-fitted by an exponential function.}
\label{fig15}
\end{center}
\end{figure}

\begin{figure}[ht]
\begin{center}
\includegraphics[height=8.0cm]{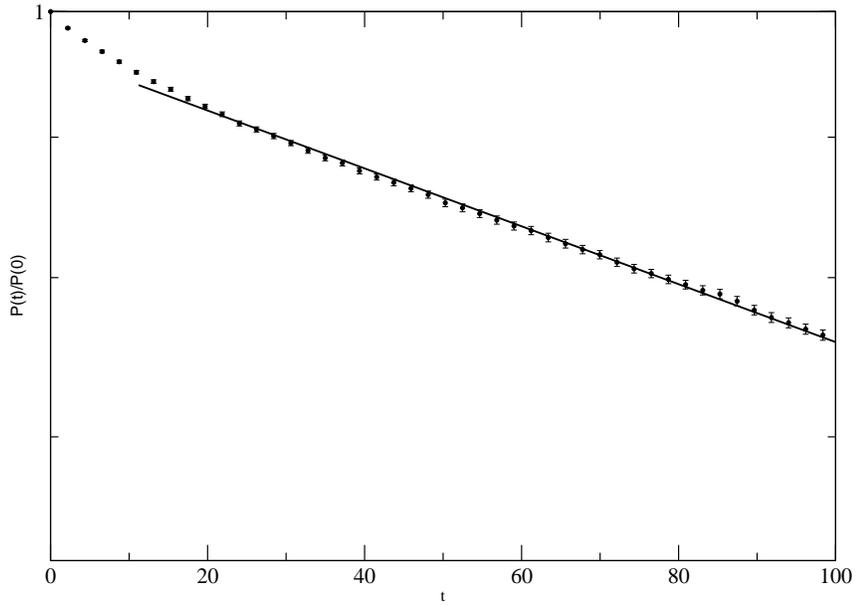}
\caption{Stress relaxation as function of time for the same system of figure (\ref{fig15}).}
\label{fig16}
\end{center}
\end{figure}

Our computational experiments give a well defined exponential behaviour, both in the system energy as in stress relaxation.  After a transient, the system energy relaxation is well fitted by a straight line in a semi-logarithmic plot for a long time, shown in figure (\ref{fig15}).  The same behaviour was found in the stress relaxation, figure (\ref{fig16}). In fact, since the system energy changes moderately in this kind of lattice, one needs a long time to see any difference in its evolution.

\begin{figure}[ht]
\begin{center}
\includegraphics[height=7cm]{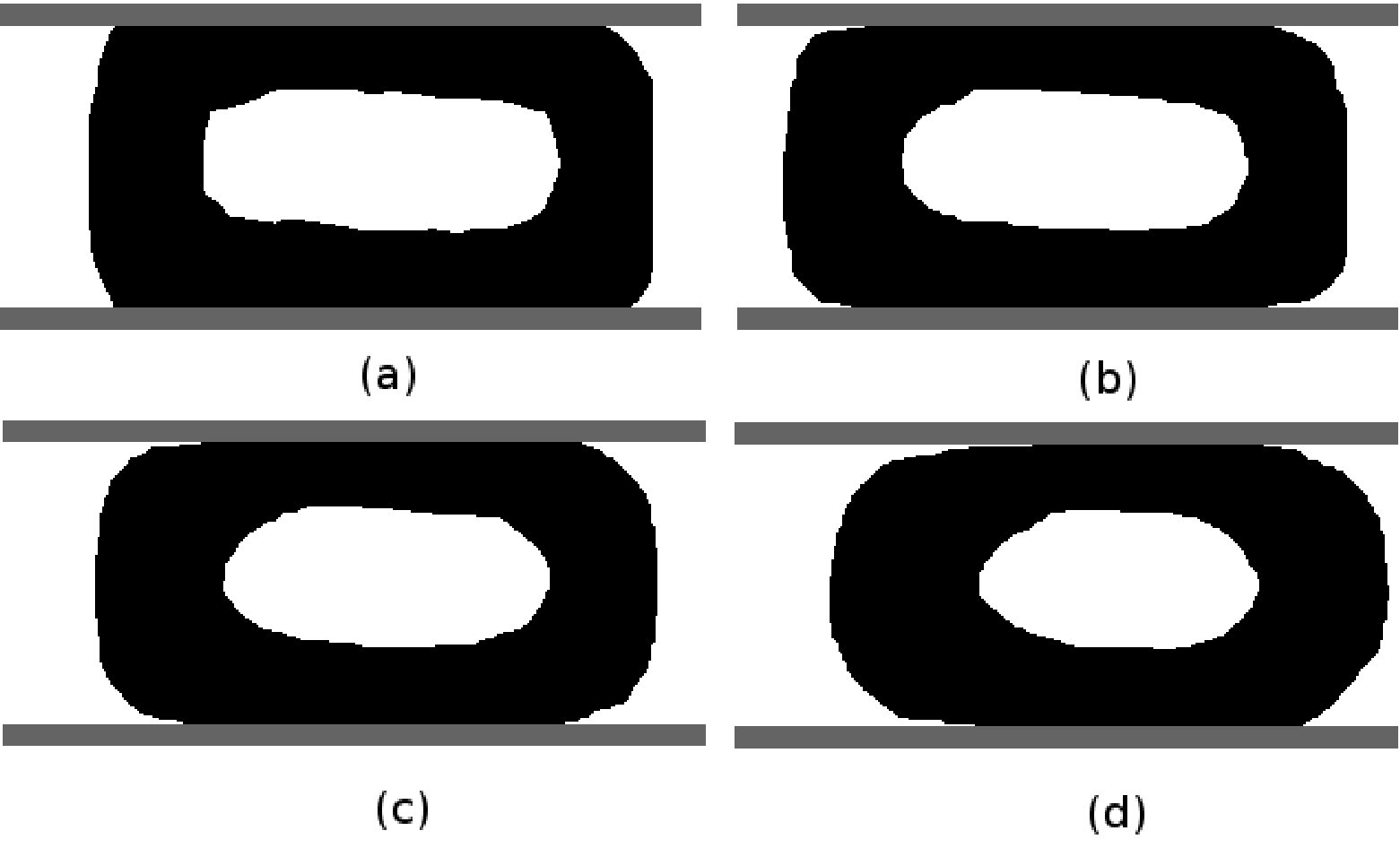}
\caption{Some steps of evolution from a non-fractal surface of mass $M = 23029$. (a) Initial lattice created from a graph editor, (b) lattice after $t = 50 M$ steps, (c) lattice after $t = 100M$ steps and (d) lattice after $t = 200M$ steps.}
\label{fig17}
\end{center}
\end{figure}

Figure (\ref{fig17}) shows the evolution of the non-fractal lattice. When the time goes by, the surface becomes more and more spherical, i.e., its geometrical shape is equilibrated by the relaxation process. Actually, it does not become completely spherical since it has a geometrical limit represented by the plates.

\section{Conclusion}

In summary, we have shown that fractal structures like Sierpinski carpet, Sierpinski gasket and percolation cluster have slow relaxation. In particular we simulate the system energy and the stress relaxation to these fractal lattices. Our computational data are well fitted by a stretched exponential function. Moreover, we have also confirmed that a non-fractal structure presents an exponential relaxation. Based on these results, we suggest that non-exponential behaviour can be related to the fractal structure of the initial crumpled surface.

Our results have also indicated that for different values of the fractal dimension we have slow relaxation with different stretched $\beta$ exponents. Such results have not been studied experimentally up to now but, certainly, other studies could be done to search for a possible relationship between the fractal dimension of the initial system and the $\beta$ relaxation exponent.

\section{Acknowledgments}

We wish to thank Professor D. Stauffer for helpful criticism and encouragement. This work was supported by CNPq (Brazilian government agency) and PRONEX-CNPq-FAPERJ/171.168-2003.

\newpage
\begin{center} 
 \textbf{reply to the referees}  
\end{center}

\newpage



\begin{thebibliography}{00}
\bibitem{13} Detailed description of the algorithm can be found in Numerical Recipes in C, Chapter 15.5: Nonlinear models.
\end{thebibliography}
\end{document}